\documentclass{iau} 
\usepackage{graphicx}

\newcommand{\oversim}[2]{\protect{\mbox{\lower0.5ex\vbox{%
  \baselineskip=0pt\lineskip=0.2ex
  \ialign{$\mathsurround=0pt #1\hfil##\hfil$\crcr#2\crcr\sim\crcr}}}}}
\newcommand{\simgreat}{\mbox{$\,\mathrel{\mathpalette\oversim>}\,$}} 
\newcommand{\simless} {\mbox{$\,\mathrel{\mathpalette\oversim<}\,$}} 

\title[The systematically varying IMF] 
{The systematically varying stellar IMF}

\author[Pavel Kroupa]   
{Pavel Kroupa}

\affiliation{Helmholtz-Institut f\"ur Strahlen- und Kernphysik,
  Universit\"at Bonn\\ 
Nussallee 14-16, 53115 Bonn, Germany \\[\affilskip]
Charles University in Prague, 
Faculty of Mathematics and Physics, 
Astronomical Institute\\
V  Hole\v{s}ovi\v{c}k\'ach 2,
CZ-18000 Praha,
Czech Republic
\\email: {\tt pkroupa@uni-bonn.de, kroupa@sirrah.troja.mff.cuni.cz}}

\pubyear{2019}
\volume{351}  
\jname{Star Clusters: From the Milky Way to the Early Universe}
\editors{A. Bragaglia, M.B. Davies, A. Sills \& E. Vesperini, eds.}
\begin{document}

\maketitle

\begin{abstract}
  Some ultra-compact dwarf galaxies have large dynamical mass to light
  ($M/L$) ratios and also appear to contain an overabundance of LMXB
  sources, and some Milky Way globular clusters have a low
  concentration and appear to have a deficit of low-mass stars. These
  observations can be explained if the stellar IMF becomes
  increasingly top-heavy with decreasing metallicity and increasing
  gas density of the forming object. The thus constrained stellar IMF
  then accounts for the observed trend of metallicity and $M/L$ ratio
  found amongst M31 globular star clusters. It also accounts for the
  overall shift of the observationally deduced galaxy-wide IMF from
  top-light to top-heavy with increasing star formation rate amongst
  galaxies. If the IMF varies similarly to deduced here, then
  extremely young very massive star-burst clusters observed at a high
  redshift would appear quasar-like (\cite{Jerab+17}) .
\end{abstract}

\firstsection 
\section{Introduction}

The stellar IMF is valid for a simple stellar population as emerges,
on a time-scale of about one Myr, from a molecular cloud core (i.e. on
its dynamical collapse time scale) in which forms an embedded cluster
on a scale of a pc or less containing dozens to many millions of
stars. The galaxy-wide IMF, gwIMF, in contrast is the composite IMF
resulting from the addition of the individual IMFs forming throughout
the system.

The above is summarised in this contribution, with the observationally
constrained gwIMF being used as a boundary condition on the
independently deduced variation of the IMF. Definitions:
\begin{itemize}

\item The {\it stellar IMF} (IMF), $\xi(m) = dN/dm$, where $dN$ is the
  infinitesimal number of stars born together (in one embedded
  cluster) with individual masses in the interval $m$ to $m+dm$. The
  {\it canonical IMF} has the shape $\xi(m)\propto m^{-\alpha_i}$,
  with $\alpha_2 \approx 2.3, 0.5\,M_\odot < m < m_{\rm max}$ being
  the Salpeter power-law index and
  $\alpha_1\approx 1.3, m<0.5\,M_\odot$. Here, $m_{\rm max}$ is the
  mass of the most massive star forming in an embedded cluster with a
  total stellar mass of $M_{\rm ecl}$ (\cite{Weidner+13}).  Since
  stars typically form as binary systems (\cite{GK05}), the molecular
  cloud filament density variations which fragment to stars
  (\cite{Andre+10}) are related to the initial system mass function
  (eq.~4-57 in \cite{Kroupa+13}).

\item The {\it galaxy-wide IMF} (gwIMF) is the sum of all IMFs over a
  whole galaxy (\cite{Yan+17,Jerab+18, Hopkins18}). The calculation of
  the gwIMF is performed, in the integrated galaxy IMF (IGIMF) theory,
  by integrating over all embedded clusters formed in the galaxy over
  a period of about~10~Myr. The 10~Myr time-scale (see
  \cite{Schulz+15, Yan+17, Jerab+18} for discussions) is given by the
  time-scale for local collapse of the interstellar medium (ISM), the
  thickness of the gas disk being about 100~pc and the velocity
  dispersion of the warm ISM being about 10~pc/Myr. In general, the
  shapes, gwIMF$\ne$IMF, because the $m_{\rm max}-M_{\rm ecl}$
  relation (\cite{KM11, KM12, Ramirez+16,Stephens+17,OK18}) implies
  low-mass embedded clusters to not have massive stars.  Within the
  IGIMF theory, the gwIMF is only comparable to the IMF in a galaxy
  with a star-formation rate (SFR) $SFR\approx 1\,M_\odot$/yr
  occurring at about solar metallicity (\cite{Yan+17, Jerab+18,
    Zonoozi+19}). The IGIMF varies systematically with the star
  formation rate and metallicity of the galaxy because the IMF varies
  on the molecular-cloud-core scale. This variation of the IMF and of
  the gwIMF carries implications for extragalactic astrophysics and
  for cosmology.

\end{itemize}

\section{The IMF from solar-neighbourhood star counts}

A vast amount of effort has been conducted to constrain the stellar
IMF from the nearby stars, starting with \cite{Salpeter55} through to
\cite{MS79} and \cite{Scalo86}. That the internal structure of stars
changes around $0.3\,M_\odot$ (less massive stars being fully
convective) led to the realisation (\cite{KTG90,KT97}) that the
stellar luminosity function ($\Psi(M_X)=dN/dM_X$, where $M_X$ is the
absolute stellar magnitude in the photometric $X$-band, e.g. $X=V$)
has a pronounced maximum at $M_V \approx 11.5, M_{\rm I} \approx
8.5$. Together with corrections of the star-counts for unresolved
binary stars, biases in photometric- and trigonometric parallaxes as
well as the age, metallicity and spatial distribution of low-mass
stars of different masses established the canonical IMF for
$m\simless 1\,M_\odot$ (\cite{KTG93}).  This form, corrected for the
first time for unresolved binary systems, is virtually identical to
the confirmation (subject to the caveat that corrections for
unresolved binary stars need to be applied properly) by
\cite{Chabrier03, Bochanski+10}, who however, used a log-normal form
for $\xi(m)$ for $m<1\,M_\odot$.  For $m\simgreat 1\,M_\odot$,
significant corrections for short stellar life times and for the
different spatial distribution of the massive stars compared to
low-mass stars need to be taken into account.  The Milky-Way (MW)
gwIMF has $\alpha_3 \approx 2.7$ (\cite{Scalo86}), while more recent
determinations yield $\alpha_3 \approx 2$ (\cite{Mor+19, Zonoozi+19}).

Important is to realise that this solar-neighbourhood IMF is in fact
the MW gwIMF, sampled from the stars within the solar
neighbourhood. The true form of the MW gwIMF remains disputable,
because the stellar ensembles used to constrain it differ: stars with
$m\simless 0.5\,M_\odot$ come from a region around the Sun spanning
only a few~pc; stars with $m\approx 1\,M_\odot$ extend the ensemble to
about 20~pc while more massive stars are taken from regions of the
Galactic disk extending to kpc distances. Thus, differences in the
velocity distribution function (and thus vertical scale height) and
the time variation of the past local SFR have an impact on the form of
the local gwIMF deduced from the star counts (\cite{ES06}).

\section{The IMF from embedded, open and globular star clusters and from
  ultra-compact dwarf galaxies}
\label{sec:IMF}

Direct star counts in nearby embedded and young clusters have been
taken to imply an invariant IMF which is consistent with the canonical
form (\cite{Bastian+10, Offner+14}).  But perhaps the most dramatic
example of a non-canonical stellar population is the observationally
deduced IMF near the Galactic centre where stars appear to have formed
in the past few~Myr with $\alpha_3\approx 0.45$ and very few if any
detected low-mass stars (\cite{Bartko+10}).  This suggests a strong
variation of the shape of the IMF with the physical conditions of the
star forming gas, whereby shear may be playing a significant role in
this particular case.  The dynamical mass-to-light ratio, $M/L$, of
some ultra-compact dwarf galaxies (UCDs, ``Hilker objects'',
\cite{Hilker+99}) can be represented by an initial stellar population
with a top-heavy IMF which depends on the birth density of the UCD
(\cite{Dab+09}). Some UCDs have an oversurplus of low-mass X-ray
binaries, which can be explained by the same dependency of the IMF
(\cite{Dab+12}). A very young UCD with a top-heavy IMF expands to
become a present-day UCD (\cite{Dab+10}). Some globular clusters (GCs)
have a deficit of low-mass stars and at the same time a low
concentration (\cite{deMarchi+07}). While inconsistent with
energy-equipartition-driven cluster evolution, this trend can be
explained by an IMF which becomes increasingly top-heavy with
decreasing metallicity and increasing density (\cite{Marks+12}). A
non-trivial outcome of this work is that the UCD and GC calculations
lead to the same IMF variation. This is non-trivial because the data
sets are completely different involving different physical processes,
yet yield a consistent result. The GCs in Andromeda show a $M/L$ trend
with metallicity which is explainable by this same IMF variation
(\cite{ZHK16, Haghi+17}). Local star-burst clusters are efficient
ejectors of their massive stars (\cite{Oh+15, OK16}). When
statistically adding them back into an observed very young cluster,
such as R136 in the Large Magellanic Cloud, the IMF of R136 is implied
to have been top-heavy (\cite{BK12}). Direct star-counts have
confirmed this (\cite{Schneider+18}). Direct star-counts also support
the IMF of low-metallicity very young clusters to be top-heavy
(\cite{Kalari+18}).

In summary, the evidence for a systematic IMF variation has become
very significant. While the detailed form of the variation remains
unclear to some degree, the particular formulation available
(\cite{Marks+12}) appears to largely embrace this variation. As a note
for completeness, MW star counts for different populations with
different metallicity have been noted to suggest the IMF to be bottom
heavy at super-solar metallicity and bottom-light at low-metallicity
(see \cite{Marks+12} and references therein).

\section{Consistency with theoretical expectations}

Until about~2010 the lack of evidence for a variable IMF was
disconcerting (see discussion in \cite{Kroupa+13}). The evidence which
started forthcoming thereafter (Sec.~\ref{sec:IMF}) became possible
through improved resolved stellar population data and improved
theoretical understanding of the stellar and dynamical evolution of
these populations.

The theoretical expectation for a variable IMF has been rigorous and
time-lasting, because two main broad but entirely independent
arguments lead to the same expectation of an IMF which should shift to
top-heavy with decreasing metallicity and increasing density. The one
argument rests on the Jeans mass instability in a molecular cloud
(\cite{Larson98}) while the other argument rests on self-regulation of
the accretion flow onto a forming star (\cite{AF96,AL96}). Metallicity
plays a decisive role in so far as it regulates the cooling through
line-emission of the collapsing gas cloud and the coupling of the
stellar photons to the accretion flow, as is indeed supported
observationally (\cite{deMarchi+17}).  Auxiliary arguments pertain to
the coagulation of proto-stars at high densities of the
embedded-cluster forming cloud core (\cite{Dib07}) and heating of the
cloud core through supernova type~II generated cosmic rays in
star-burst regions (\cite{Pap10}).

The recent developments in IMF studies (Sec.~\ref{sec:IMF} and this
section) have thus nicely shown broad convergence of empirical and
theoretical results.

\section{Consistency with galaxy-wide stellar populations}
\label{sec:gwIMF}

The IMF variation suggested above ought to be evident in a
corresponding variation of the gwIMF, since, trivially, the young
population in a galaxy is composed of the sum over all its embedded
clusters.

The G-dwarf problem (too few low-metallicity G-dwarfs found locally)
has been used to argue that the gwIMF must have been top-heavy in past
cosmological times (e.g. \cite{Dave08}).  The emission of H$\alpha$
light is a direct tracer of the young ionising massive-stellar content
of a galaxy with stellar life-times up to about 50~Myr, while the
brightness in the UV spectral bands assesses the intermediate-mass
stellar population with stellar life-times of up to about~300~Myr
(\cite{Kennicutt98, PWK09, Jerab+18}). Broad-band optical colours also
allow an assessment of the intermediate to low-mass stellar
population. A combination of these tracers thus allows, under
appropriate consideration of photon leakage and dust obscuration and
scattering, information to be gathered on the shape of the gwIMF.
Surveys of local-volume star-forming dwarf galaxies have shown these
to have a decreasing H$\alpha$/UV flux ratio with decreasing SFR
(implying an increasingly top-light gwIMF with decreasing SFR,
\cite{Lee+09}).  Star-counts in the $\approx 4\,$Mpc distant dwarf
galaxy DDO~154 with a $SFR\approx 10^{-3}\,M_\odot$/yr have uncovered
a deficit of massive stars (\cite{Watts+18}), consistent with the
above survey.  Surveys of disk galaxies with
$SFR \simgreat 1\,M_\odot$/yr imply an increasingly top-heavy gwIMF
with increasing SFR (\cite{Gunn+11}). Independently-performed surveys
of star-forming galaxies come to the same conclusion (\cite{HG08,
  Meurer+09}). This observationally inferred general trend of the
gwIMF becoming increasingly top-heavy with increasing SFR is
consistent with the early deduction based on $\alpha$- and Fe-element
abundances in elliptical galaxies which formed with SFRs as high as a
few~$10^3\,M_\odot$/yr (\cite{Matteucci94}).  A potentially powerfull
tool to assess the shape of the gwIMF in high-redshift star-bursting
galaxies is the $^{13}{\rm C}/^{18}{\rm O}$ isotope abundance ratio in
the cold molecular gas. Probing it via the rotational transitions of
the $^{13}{\rm CO}$ and ${\rm C}^{18}{\rm O}$ isotopologues provides a
sensitive measure of the slope of the gwIMF at high stellar
masses. Using this method leads to the result that the gwIMF becomes
increasingly top heavy with increasing SFR (\cite{Zhang+18}).

This documented variation of the gwIMF from the least-massive
star-forming dwarf galaxies to the most massive now dormant elliptical
galaxies is well described by the IGIMF theory (\cite{Yan+17,
  Jerab+18}).

\section{Final comments}

The evidence for the IMF and, by implication, also for the gwIMF to
vary appears to be rather robust. Various independent lines of
argument, also based on independent data, point to the IMF being more
top-heavy at low metallicity, high density, and thus to have been
top-heavy in the early Universe. This is well consistent with
theoretical expectations and with galaxy-wide stellar populations. An
explicit mathematical formulation of this dependency is available
(\cite{Marks+12}, \cite{Jerab+18}).  Observations at high redshift and
thus in extreme environments will allow this formulation to be
updated, but any changes must always retain consistency with the
observational constraints in the MW. The current prediction, based on
an extrapolation to extreme conditions, is that cosmologically early
star-burst clusters (proto-UCDs) would appear quasar-like
(\cite{Jerab+17}). It may thus be that some of the very high-red shift
quasars may not be accreting super massive black holes, but may rather
be a prior stage, namely a few-Myr old hyper-star-burst clusters.


\begin{thebibliography}{}

\bibitem[Adams \& Fatuzzo 1996]{AF96} Adams, F.~C., \& Fatuzzo, M.\
  1996, \textit{ApJ}, 464, 256

\bibitem[Adams \& Laughlin 1996]{AL96} Adams, F.~C., \& Laughlin, G.\
  1996, \textit{ApJ}, 468, 586

\bibitem[Andr{\'e} et al. 2010]{Andre+10} Andr{\'e}, P., Men'shchikov,
  A., Bontemps, S., et al.\ 2010, \textit{A\&A}, 518, L102

\bibitem[Banerjee \& Kroupa 2012]{BK12} Banerjee, S., \& Kroupa, P.\ 2012, \textit{A\&A}, 547, A23

\bibitem[Bartko et al. 2010]{Bartko+10} Bartko, H., Martins, F.,
  Trippe, S., et al.\ 2010, \textit{ApJ}, 708, 834

\bibitem[Bastian et al. 2010]{Bastian+10} Bastian, N., Covey, K.~R.,
  \& Meyer, M.~R.\ 2010, \textit{ARAA}, 48, 339

\bibitem[Bochanski et al. 2010]{Bochanski+10} Bochanski, J.~J.,
  Hawley, S.~L., Covey, K.~R., et al.\ 2010, \textit{AJ}, 139, 2679

\bibitem[Chabrier 2003]{Chabrier03} Chabrier, G.\ 2003,
 \textit{PASP}, 115, 763

\bibitem[Dabringhausen et al. 2009]{Dab+09} Dabringhausen, J., Kroupa,
  P., \& Baumgardt, H.\ 2009, \textit{MNRAS}, 394, 1529

\bibitem[Dabringhausen et al. 2010]{Dab+10} Dabringhausen, J.,
  Fellhauer, M., \& Kroupa, P.\ 2010, \textit{MNRAS}, 403, 1054

\bibitem[Dabringhausen et al. 2012]{Dab+12} Dabringhausen, J., Kroupa,
  P., Pflamm-Altenburg, J., et al.\ 2012, \textit{ApJ}, 747, 72

\bibitem[Dav{\'e} 2008]{Dave08} Dav{\'e}, R.\ 2008, \textit{MNRAS},
  385, 147

\bibitem[De Marchi et al. 2007]{deMarchi+07} De Marchi, G., Paresce, F., \& Pulone, L.\ 2007, \textit{ApJL}, 656, L65

\bibitem[De Marchi et al. 2017]{deMarchi+17} De Marchi, G., Panagia, N., \& Beccari, G.\ 2017, \textit{ApJ}, 846, 110

\bibitem[Dib et al. 2007]{Dib07} Dib, S., Kim, J., \& Shadmehri, M.\ 2007, \textit{MNRAS}, 381, L40


\bibitem[Elmegreen \& Scalo 2006]{ES06} Elmegreen, B.~G., \& Scalo,
  J.\ 2006, \textit{ApJ}, 636, 149

\bibitem[Goodwin \& Kroupa 2005]{GK05} Goodwin, S.~P., \& Kroupa, P.\
  2005, \textit{A\&A}, 439, 565

\bibitem[Gunawardhana et al. 2011]{Gunn+11} Gunawardhana, M.~L.~P.,
  Hopkins, A.~M., Sharp, R.~G., et al.\ 2011, \textit{MNRAS}, 415,
  1647

\bibitem[Haghi et al. 2017]{Haghi+17} Haghi, H., Khalaj, P., Hasani
  Zonoozi, A., et al.\ 2017, \textit{ApJ}, 839, 60

\bibitem[Hilker et al. 1999]{Hilker+99} Hilker, M., Infante, L.,
  Vieira, G., et al.\ 1999, \textit{A\&AS}, 134, 75

\bibitem[Hopkins 2018]{Hopkins18} Hopkins, A.~M.\ 2018, \textit{PASA},
  35, 39

\bibitem[Hoversten, \& Glazebrook 2008]{HG08} Hoversten, E.~A., \&
  Glazebrook, K.\ 2008, \textit{ApJ}, 675, 163

\bibitem[Jerabkova et al. 2017]{Jerab+17} Jerabkova, T., Kroupa, P.,
  Dabringhausen, J., et al.\ 2017, \textit{A\&A}, 608, A53

\bibitem[Jerabkova et al. 2018]{Jerab+18} Jerabkova, T., Hasani
  Zonoozi, A., Kroupa, P., et al.\ 2018, \textit{A\&A}, 620, A39

\bibitem[Kalari et al. 2018]{Kalari+18} Kalari, V.~M., Carraro, G.,
  Evans, C.~J., et al.\ 2018, \textit{ApJ}, 857, 132


\bibitem[Kennicutt 1998]{Kennicutt98} Kennicutt, R.~C.\ 1998,
  \textit{ARAA}, 36, 189


\bibitem[Kirk \& Myers 2011]{KM11} Kirk, H., \&
  Myers, P.~C.\ 2011, \textit{ApJ}, 727, 64

\bibitem[Kirk \& Myers 2012]{KM12} Kirk, H., \& Myers, P.~C.\ 2012, \textit{ApJ}, 745, 131


\bibitem[Kroupa et al. 1990]{KTG90} Kroupa, P., Tout, C.~A., \&
  Gilmore, G.\ 1990, \textit{MNRAS}, 244, 76

\bibitem[Kroupa et al. 1993]{KTG93} Kroupa, P., Tout, C.~A., \&
  Gilmore, G.\ 1993, \textit{MNRAS}, 262, 545

\bibitem[Kroupa \& Tout 1997]{KT97} Kroupa, P., \& Tout, C.~A.\ 1997,
  \textit{MNRAS}, 287, 402

\bibitem[Kroupa et al. 2013]{Kroupa+13} Kroupa, P., Weidner, C.,
  Pflamm-Altenburg, J., et al.\ 2013, \textit{Planets, Stars and
    Stellar Systems}. Volume 5: Galactic Structure and Stellar
  Populations, 115


\bibitem[Larson 1998]{Larson98} Larson, R.~B.\ 1998, \textit{MNRAS}, 301, 569

\bibitem[Lee et al. 2009]{Lee+09} Lee, J.~C., Gil de Paz, A.,
  Tremonti, C., et al.\ 2009, \textit{ApJ}, 706, 599

\bibitem[Marks et al. 2012]{Marks+12} Marks, M., Kroupa, P.,
  Dabringhausen, J., et al.\ 2012, \textit{MNRAS}, 422, 2246

\bibitem[Matteucci 1994]{Matteucci94} Matteucci, F.\ 1994,
  \textit{A\&A}, 288, 57


\bibitem[Meurer et al. 2009]{Meurer+09} Meurer, G.~R., Wong, O.~I.,
  Kim, J.~H., et al.\ 2009, \textit{ApJ}, 695, 765


\bibitem[Miller \& Scalo 1979]{MS79} Miller, G.~E., \& Scalo,
  J.~M.\ 1979, \textit{ApJS}, 41, 513

\bibitem[Mor et al. 2019]{Mor+19} Mor, R., Robin, A.~C., Figueras, F.,
  et al.\ 2019, \textit{A\&A}, 624, L1

\bibitem[Offner et al. 2014]{Offner+14} Offner, S.~S.~R., Clark,
  P.~C., Hennebelle, P., et al.\ 2014, in \textit{Protostars and Planets VI}, 53


\bibitem[Oh et al. 2015]{Oh+15} Oh, S., Kroupa, P., \& Pflamm-Altenburg, J.\ 2015,
  \textit{ApJ}, 805, 92

\bibitem[Oh \& Kroupa 2016]{OK16} Oh, S., \& Kroupa, P.\ 2016,
  \textit{A\&A}, 590, A107

\bibitem[Oh \& Kroupa 2018]{OK18} Oh, S., \& Kroupa, P.\ 2018,
  \textit{MNRAS}, 481, 153

\bibitem[Papadopoulos 2010]{Pap10} Papadopoulos, P.~P.\ 2010,
  \textit{ApJ}, 720, 226

\bibitem[Pflamm-Altenburg et al. 2009]{PWK09} Pflamm-Altenburg, J.,
  Weidner, C., \& Kroupa, P.\ 2009, \textit{MNRAS}, 395, 394

\bibitem[Ram{\'{\i}}rez Alegr{\'{\i}}a et al. 2016]{Ramirez+16}
  Ram{\'{\i}}rez Alegr{\'{\i}}a, S., Borissova, J., Chen{\'e}, A.-N.,
  et al.\ 2016, \textit{A\&A}, 588, A40

\bibitem[Salpeter 1955]{Salpeter55} Salpeter, E.~E.\ 1955, \textit{ApJ}, 121, 161

\bibitem[Scalo 1986]{Scalo86} Scalo, J.~M.\ 1986,
  \textit{Fundamentals of Cosmic Physics}, 11, 1

\bibitem[Schneider et al. 2018]{Schneider+18} Schneider, F.~R.~N.,
  Sana, H., Evans, C.~J., et al.\ 2018, \textit{Science}, 359, 69

\bibitem[Schulz et al. 2015]{Schulz+15} Schulz, C., Pflamm-Altenburg,
  J., \& Kroupa, P.\ 2015, \textit{A\&A}, 582, A93

\bibitem[Stephens et al. 2017]{Stephens+17} Stephens, I.~W.,
  Gouliermis, D., Looney, L.~W., et al.\ 2017, \textit{ApJ}, 834, 94

\bibitem[Watts et al. 2018]{Watts+18} Watts, A.~B., Meurer, G.~R.,
  Lagos, C.~D.~P., et al.\ 2018, \textit{MNRAS}, 477, 5554

\bibitem[Weidner et al. 2013]{Weidner+13} Weidner, C., Kroupa, P., \&
  Pflamm-Altenburg, J.\ 2013, \textit{MNRAS}, 434, 84


\bibitem[Yan et al. 2017]{Yan+17} Yan, Z., Jerabkova,
  T., \& Kroupa, P.\ 2017, \textit{A\&A}, 607, A126

\bibitem[Zhang et al. 2018]{Zhang+18} Zhang, Z.-Y., Romano, D.,
  Ivison, R.~J., et al.\ 2018, \textit{Nature}, 558, 260

\bibitem[Zonoozi et al. 2016]{ZHK16} Zonoozi, A.~H., Haghi, H., \&
  Kroupa, P.\ 2016, \textit{ApJ}, 826, 89

\bibitem[Zonoozi et al. 2019]{Zonoozi+19} Zonoozi, A.~H., Mahani, H.,
  \& Kroupa, P.\ 2019, \textit{MNRAS}, 483, 46

\end{thebibliography}
\end{document}